# Public EV Charging Choices: How Users Trade Off Time, Price, and Renewable Energy

Delong Du
Delong.Du@uni-siegen.de
University of Siegen
Siegen, Germany

Apostolos K. Vavouris
apostolos.vavouris@strath.ac.uk
University of Strathclyde
Glasgow, United Kingdom

Omid Veisi
omid.veisi@student.uni-siegen.de
University of Siegen
Siegen, Germany

Lu Jin
lu.jin@fit.fraunhofer.de
Fraunhofer FIT
Sankt Augustin, Germany

Gunnar Stevens
gunnar.stevens@uni-siegen.de
University of Siegen
Siegen, Germany

Lina Stankovic
lina.stankovic@strath.ac.uk
University of Strathclyde
Glasgow, United Kingdom

Vladimir Stankovic
vladimir.stankovic@strath.ac.uk
University of Strathclyde
Glasgow, United Kingdom

Alexander Boden
alexander.boden@fit.fraunhofer.de
Fraunhofer FIT
Sankt Augustin, Germany

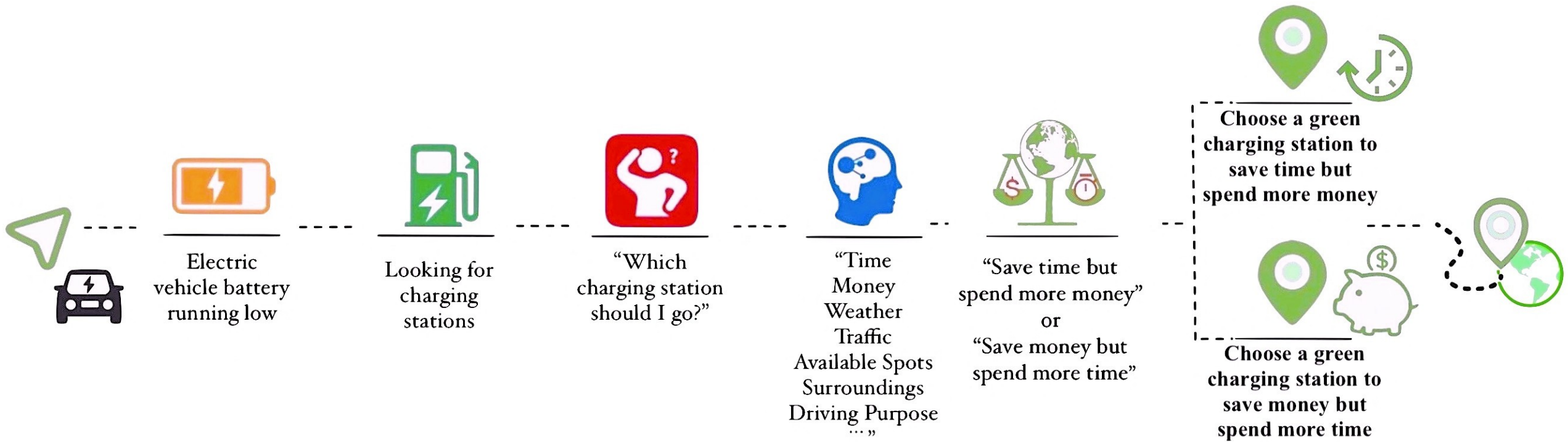

**Figure 1: Design synthesis of the considerations participants described when evaluating a renewable-energy-aware charging recommendation.**

## ABSTRACT

The carbon intensity of electric-vehicle (EV) charging varies over time and place, yet EV charging recommender systems and eco-routing interfaces rarely make this variation actionable for drivers. We investigate how renewable-energy information interacts with two attributes that routinely shape public-charging decisions: travel time and price. Fifty car users completed a within-subjects stated-choice study with three navigation-interface variants, and 10 EV drivers participated in semi-structured interviews. Across the three variants, the share choosing the slower option was 38%, 42%, and 66%, respectively. A paired-sample analysis found that choices differed across variants (Cochran's $Q(2) = 11.47$, $p = .003$). When the time–price trade-off was held constant, adding a renewable-energy label increased selection of the slower station from 38% to 66% (exact McNemar test, Holm-adjusted $p = .004$). Interviews nevertheless showed that renewable energy was usually a secondary consideration: participants evaluated it through situational constraints such as urgency, charging cost, traffic, charger availability, and familiarity with locations. We derive a constraint-first design rationale for renewable-energy-aware charging recommendations: filter options using context-sensitive time and cost constraints, disclose the renewable-energy signal and its uncertainty, and preserve user control rather than relying on a green default alone. Together, the results provide an empirical baseline for research on green charging recommendations, while characterizing stated choices in a small exploratory sample rather than real-world charging behavior.

# 1 INTRODUCTION

Human-computer interaction (HCI) research has long explored how interactive systems can support environmentally preferable decisions in mobility [27, 39, 41], food consumption [26, 28, 29, 55], tourism [50], and household energy use [6, 7, 46–49, 51]. Public EV charging creates a particularly consequential design problem: the lower-carbon option may be farther away, slower, more expensive, unavailable on arrival, or unfamiliar to the driver. A useful interface must therefore do more than display an environmental label; it must place that information within the practical trade-offs that determine whether a recommendation is acceptable.

EVs can reduce transport-related greenhouse-gas emissions relative to internal-combustion vehicles, although the magnitude depends on vehicle production, electricity generation, and when and where charging occurs [23, 43]. Charging navigation can make part of this variability visible by identifying stations or time windows with a higher renewable share or lower carbon intensity [38, 57]. The interface challenge is to communicate that signal without implying that electricity is physically traceable to a single source or that renewable charging has zero life-cycle emissions.

However, renewable energy, despite its clean nature, is subject to fluctuation due to variation, thus making it challenging to guarantee the availability of energy and power on demand [4, 36]. To maintain a steady renewable energy supply, solutions such as sustainable hydropower storage [53], diversifying energy sources [21], and enabling grid stability for increasing photovoltaics through grid management [5] are employed to mitigate the fluctuating nature of renewable energy sources. However, the availability of renewable energy for use differs across electricity grids in various areas and regions, influenced by the diversity of their regional electricity generation mixes [62] and the varying levels of renewable energy penetration and infrastructure [37]. As microgrids increasingly adopt distributed renewable energy generation [19, 52], regional variations significantly influence the type of renewable facilities implemented, such as solar PV and wind energy systems. These differences play a critical role in shaping the energy source distribution within the grid. With the growing differences in microgrid configurations, the phenomenon of regional variations in renewable energy availability is expected to escalate, leading to high variability even between neighboring areas. Some charging stations are off-grid, powered by distributed energy sources instead of the conventional utility grid [35].

This study uses three navigation-interface prototypes as research artifacts [71] to examine stated charging-station choices. Rather than treating sustainability as an isolated preference, we study how a renewable-energy cue interacts with time and price and use interviews to understand the situational reasoning behind participants' choices. We ask:

**RQ1**: How do stated charging-station choices change across interface scenarios that vary travel time, price, and renewable-energy information?

**RQ2**: What rationales and contextual factors do EV drivers report when evaluating renewable-energy-aware charging recommendations?

We reviewed Tesla's in-car navigation interface and simplified its visual structure for the study. Figure 2 shows the three variants used to compare the roles of time, price, and renewable-energy information.

- **Variant 1 (time vs price):** a faster, more expensive option (X) versus a slower, cheaper option (Y).
- **Variant 2 (time vs renewable energy):** equally priced options, with Y slower and labeled as renewable-powered.
- **Variant 3 (time vs price plus renewable energy):** X remained faster and more expensive, while Y was slower, cheaper, and labeled as renewable-powered.

The paper makes three reusable contributions. First, it reports an interpretable empirical contrast: with the same time–price structure, adding the renewable-energy cue was associated with a 28-percentage-point increase in selection of the slower station (from 38% to 66%), with 16 participants moving from X to Y and two moving in the opposite direction. Second, the paired analysis uses each participant as their own control and reports all response-pattern counts, making the corrected result reproducible from the published aggregate record. Third, the interviews motivate a constraint-first design principle for green charging recommendations: satisfy practical time, price, compatibility, and reliability constraints before using renewable-energy information to rank the feasible options, while making the environmental signal transparent and revisable by the user.

# 2 RELATED WORK

This section discusses current research on regional variations of renewable energy availability, followed by prior work on navigation systems for EV charging stations and EV users' experience with charging stations.

## 2.1 Regional Variation of Renewable Energy Availability

Renewable energy sources are being promoted worldwide, with the International Energy Agency predicting a stabilization in electricity generation emissions from 2023 to 2025 with 42 countries from Asia, Africa, North and South America, Europe, and Australia [21]. In the United States, the share of renewable energy in the electric power sector increased from 9% in 2008 to 17% in 2018 [24]. In 2022, Sweden led the EU with 66.0% of its energy from renewables, primarily hydro, wind, biofuels, and heat pumps; Finland followed at 47.9%, using hydro, wind, and biofuels; Denmark (41.6%) used wind and biofuels [12].

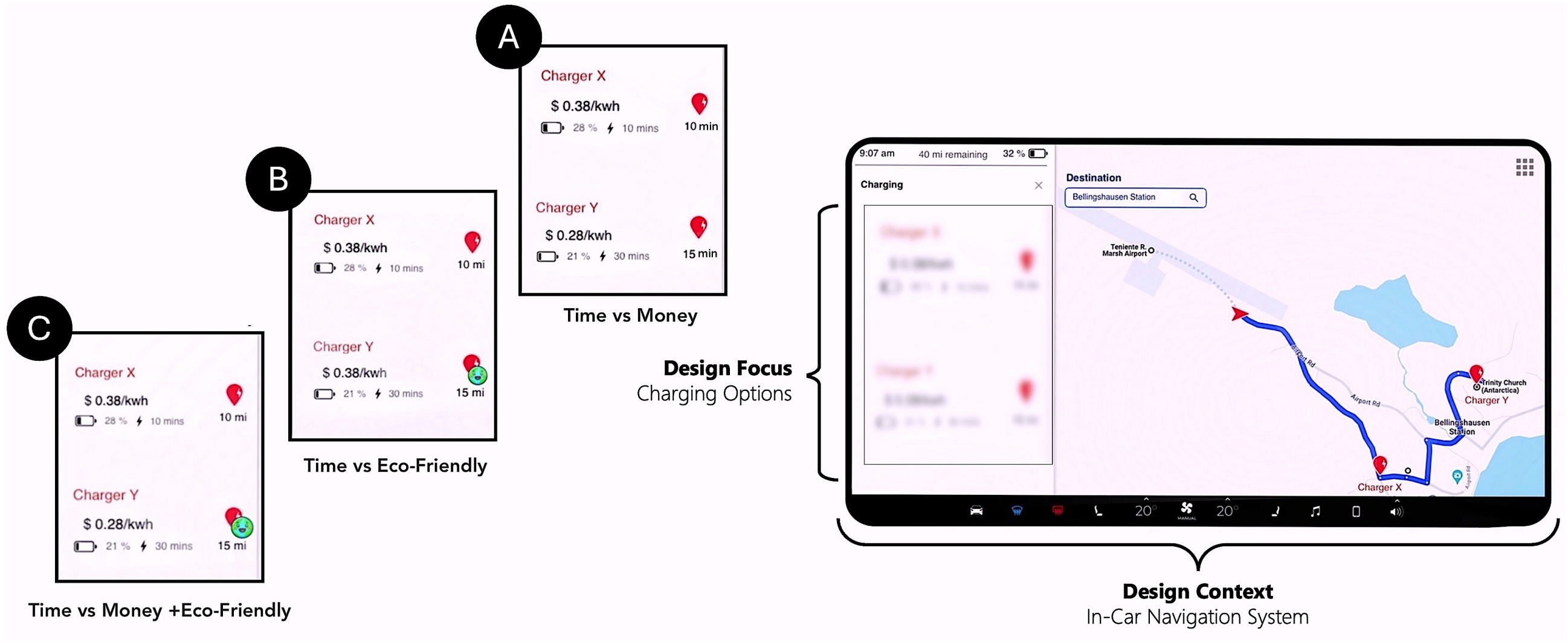

**Figure 2: The three within-subjects navigation-interface variants. Variant 1 presents a time–price trade-off; Variant 2 keeps price equal and marks the slower option as renewable-powered; Variant 3 restores the time–price trade-off and also marks the slower, cheaper option as renewable-powered.**

To reduce carbon emissions, EVs can ideally be charged with renewable sources rather than non-renewable energy [23, 62]. The carbon footprint of an EV is correlated with varying patterns of charging of renewable energy availability, as well as with the country and the area within a country where the charger is located [62]. However, varying levels of renewable energy availability are observed, notably experiencing spatial and temporal heterogeneity. Various factors, such as investment efforts, consumption intensity, and generation efficiency, influence the mechanisms for renewable generation and infrastructure penetration[68]. These can be influenced by the diversity of weather and geographical conditions and the renewable energy availability in different areas and regions [67]. Although electricity networks are generally interconnected on a regional, national, and international level, the energy produced is first consumed locally and then exported to other regions, as transportation of energy over long distances carries high costs due to the inherent complexity of constructing and maintaining the required high power lines, and due to losses in the transportation and distribution network. These losses (circa 8% of the total electricity produced) are responsible for approximately 1Gt of CO2 emissions per year [20]. In the USA, in 2022, more than 0.8 quadrillion British thermal units were lost due to the transmission and distribution of electricity [61]. In China, smart grids experience a regional imbalance between power resources that requires long transmission distances from 1000-3000 km [69]. The renewable energy availability of each microgrid can vary greatly between microgrids of different areas. In the United Kingdom, across the 14 different areas of the Distribution System Operators grids, the difference in penetration of renewables (especially between North and South) leads to a significant discrepancy between the availability of renewable energy [37].

With the ever-increasing pace of penetration of small-scale distributed renewables, distribution and transmission losses are expected to decrease, as electricity will need to travel less distance between the production and consumption point [59]. Introducing distributed sources will result in a highly complex system with different generation mixtures (and carbon footprints) even at the subregional/neighborhood level, with research focusing on optimal ways to distribute this generation sources [59]. This will lead to different availability of energy, power, and pricing, even within close proximity, based on the specific characteristics of the generation mixture in each area connecting to its microgrids.

Different approaches have been proposed to harness energy from distributed sources or to recover wasted energy for EV charging. In [34], energy otherwise lost from overhead lines due to design properties and the deceleration of light rail rolling stock is used to charge EVs in parking lots located near Edinburgh Tram electricity substations. Furthermore, approaches using distributed renewables for EV charging have been proposed in the literature [45], with the authors proposing using local energy storage systems for peak saving and reducing the dependency on the utility grid. Although distributed renewables can help provide locally produced energy, these sources are characterized by high variability due to the stochastic nature of the physical phenomena responsible for renewable generation. Renewable sources generally exhibit diurnal and seasonal variations due to different solar insolation levels, wind speed, water flow, force, etc. Therefore, renewable energy's temporal and spatial availability and cost should be communicated to end users through information and communication technologies (ICT), such as the navigation system of EV charging stations, to make sustainable decisions.

## 2.2 Navigation Systems of EV Charging Stations

Traveling long distances with EVs requires recharging, making travel time optimization planning essential for practical purposes [44]. The existing state-of-the-art EV charging systems focus on integrating renewable energy availability into eco-friendly recharging navigation systems [38, 57, 66]. Two classes of EV charging navigation applications differ mainly in terms of their workflows; one is function-based, and the other is context-based, each catering to different user needs and preferences, shown in Figure 2. Function-based navigations [1] focus on presenting a straightforward and function-centric interface. Upon launching the app, users are greeted with a menu or a ranking list of functions such as finding the nearest charging station, checking station availability, filtering stations by charger type, or viewing charging prices. The usage is linear and task-orientated; users select a function and are then taken to a new screen that delves into the details of that particular function. This approach suits users who prefer direct interaction with specific features without additional contextual information. On the other hand, context-based navigations [42] take a more holistic and immersive approach. It typically opens with a map view, displaying the user's current location and nearby EV charging stations. Such navigation systems contain essential information, such as the availability of charging ports and connector types, pricing, and even user reviews. Users can interact with the map to find stations, plan routes, or get recommendations based on their current battery level and destination. The context-based approach is ideal for users who appreciate a comprehensive overview and prefer to make decisions based on a wide range of information presented within their current context.

EV Supply Equipment (EVSE) is the infrastructure facility designed to charge EVs. These stations provide electrical power to EVs to recharge their batteries. The proliferation of EV charging stations is a critical factor in the adoption and convenience of using EVs. Public charging stations can be found in various public locations, such as shopping centers, parking lots, and dedicated charging stations. The overall EV charging system includes vehicles, EV customers, and the charging station center [2]. Most navigation applications for EV charging stations include maps, navigation, and recommender systems for charging stations. The map function enables finding nearby EV charging stations and provides detailed information. The navigation function enables guiding to the nearest EV charging station; the recommender of charging stations provides a list of optimal choices and routes that connect the battery remaining with the available charging station. In addition to the main function, several other factors, such as weather and traffic conditions provided in navigation, can also influence the results of the selection route [2]. Navigation strategies can arguably be very important for EV users, even more so than for drivers of cars with combustion engines. By using these strategies, drivers can not only choose less congested routes but also charge their vehicles at less busy charging stations. This can result in a number of benefits for both the driver and the community as a whole, including better resource management, reduced range anxiety for EV users, and optimized overall costs. Without navigation, EV users may face traffic congestion, high energy consumption, long waiting times at charging stations, and ultimately higher costs [18].

## 2.3 EV Users' Experience on Charging Stations

According to ISO 9241-210:2019, User experience is defined as "a consequence of the presentation, functionality, system performance, interactive behavior, and assistive capabilities of an interactive system, both hardware and software; it is also a consequence of the user's prior experiences, attitudes, skills, habits and personality" [22]. When specifically examining EV users' experience, EV users' interactive behaviors with such navigation systems focus on their rationale for choosing charging stations powered by renewable energy. Many considerations are known to influence the decisions of EV users regarding charging stations. Factors such as budgets of time and money have been discussed to be important for selecting charging stations [8, 58, 63]. Time and money budgets are intertwined with several other factors that influence EV traveling decisions for selecting charging stations, such as EV characteristics (e.g., the type of battery) [70] or maximum AC / DC charging ratings [32], user travel time and plans [56, 64], environmental and situational context such as weather, road and traffic conditions [3, 31], as well as charging infrastructure attributes like availability of charging spots [17, 40] and surroundings [40].

With the transition towards renewable energy use for EV charging, linking electrical power systems and ICT navigation systems aims to engage EV users in providing demand response [14]. Real-time monitoring and updates on renewable energy availability require municipal administration to collect data from infrastructure, sensors, and energy sources for energy management [33]. Recent studies have demonstrated the significant potential of ICT-embedded recommender systems in optimizing the integration of EVs with renewable energy sources for changing purposes. Das et al. proposed a multi-objective techno-economic-environmental optimation approach for EV charging, which reduces energy costs and grid utilization [11]. Chen et al. introduced a blockchain-based EV incentive system to maximize renewable energy utilization by scheduling EV charging consumption [10]. Tang et al. proposed distributed routing and charging scheduling optimization problems for an Internet of Electric Vehicle (IoEV) network to minimize the money and time cost for EV users while accommodating changes in renewable energy availability [60].

Recent empirical work reinforces the importance of treating charging choice as multi-attribute and heterogeneous. A German choice experiment found that energy price and infrastructure reliability were major determinants of public-charging tariff preferences [30]. A representative stated-choice study in Israel quantified trade-offs among charging time, incentives, and green electricity and found that renewable-energy information could offset part of the discount needed to shift charging time [54]. Revealed-preference models based on actual charging sessions likewise identify distance, proximity to home, and outlet availability as important, with substantial differences among users [25]. These studies quantify infrastructure or program preferences; our HCI contribution focuses on how a renewable-energy cue is interpreted inside a navigation recommendation and why its effect depends on the accompanying time and price trade-off.

# 3 METHODOLOGY

We used an exploratory mixed-methods design comprising a within-subjects stated-choice study and follow-up interviews. The prototypes served as research artifacts for eliciting preferences rather than as a deployed recommender system [71]. The quantitative component describes how choices changed across three controlled interface scenarios; the qualitative component provides explanations and contextual detail. Because the study measured hypothetical choices, we use terms such as "choice," "association," and "preference" rather than claiming that the interfaces changed real-world charging behavior.

## 3.1 Gather User Preferences: A Within-Subjects Design Survey Approach

Fifty car users (drivers or passengers) completed all three binary choice tasks shown in Figure 2. In every task, X denoted the faster station and Y the slower station; the price relation and renewable-energy cue varied by condition. The renewable-energy option was consistently displayed as Y to reduce position switching between variants. This fixed presentation also creates a possible order or position effect, which we address as a limitation. The design supports within-person comparisons of the three stated choices [9]; it does not estimate willingness to pay or the independent utility of every attribute.

## 3.2 Quantitative Analysis

The original report summarized each condition with aggregate counts. For this revision, we used the reported participant-level response patterns (e.g., XYY) to conduct tests appropriate for repeated binary outcomes. We first applied Cochran's Q test to the three related proportions. We then used two-sided exact McNemar tests for the three pairwise comparisons and controlled the family-wise error rate with Holm's procedure. We report Holm-adjusted $p$ values. The V1–V3 comparison is the clearest contrast for the renewable-energy cue because those variants use the same time–price tradeoff; the principal difference is that Y is marked as renewable-powered in V3. The V2–V3 comparison keeps that cue on Y and changes its price advantage. V1–V2 changes both price and renewable-energy information, so it does not identify either factor by itself.

## 3.3 Understand Decision-Making Processes: An Interview Approach

After the choice tasks, we conducted in-person field interviews while driving with 10 EV users: three from rural areas, three from suburban areas, and four from urban areas. Participants reported demographic and driving information (Table 1), typical daily driving time, and estimated daily battery use. Two participants were electric-taxi drivers and reported using approximately 180% of a battery cycle per day because they recharged during the day. Other participants primarily used their vehicles for commuting or personal travel and typically charged at home about once every two days.

The interviews combined retrospective questions about participants' prior charging and navigation experiences with speculative questions grounded in the three interface variants. Participants stated their choices, explained their reasoning, and discussed conditions that might change a decision. We organized the reported rationales into recurring factors—time, price, renewable energy, traffic, charger availability, nearby amenities, and familiarity with locations—and use quotations to retain participants' own framing. The available study record does not specify a formal coding protocol or inter-coder agreement; accordingly, the qualitative findings are presented as a descriptive synthesis rather than a claim of thematic saturation.

## 3.4 Converge Mix-Method Results

We integrated the choice patterns and interview accounts in a journey map [13]. The map is an interpretive design artifact: it connects the point at which a recommendation is shown with the constraints participants said they assess before accepting it. It is not a validated behavioral model.

# 4 RESULTS

## 4.1 Stated Charging-Station Choices

Figure 3 summarizes the choices of all 50 participants. The faster option X was selected by 31 participants (62%) in V1 and 29 (58%) in V2, but by only 17 (34%) in V3. Conversely, selection of the slower option Y increased from 19 participants (38%) in V1 to 21 (42%) in V2 and 33 (66%) in V3.

- **V1, time vs price:** X = 31, Y = 19.
- **V2, time vs renewable energy:** X = 29, Y = 21.
- **V3, time vs price plus renewable energy:** X = 17, Y = 33.

The omnibus repeated-measures test indicated that the proportion choosing Y differed across variants, Cochran's $Q(2) = 11.47$, $p = .003$. Pairwise exact McNemar tests clarify where individual choices changed:

- **V1 vs V2:** 13 participants moved from X to Y and 11 moved from Y to X; the paired proportions did not differ (Holm-adjusted $p = .839$). Because both price and renewable-energy information change, this contrast should not be attributed to either factor alone.
- **V2 vs V3:** 15 participants moved from X to Y and three moved from Y to X (Holm-adjusted $p = .015$). Both variants mark Y as renewable-powered, while V3 also makes Y cheaper, consistent with price strengthening acceptance of the slower renewable option.
- **V1 vs V3:** 16 participants moved from X to Y and two moved from Y to X (Holm-adjusted $p = .004$). The time–price relation is held constant and V3 adds the renewable-energy cue to Y, making this the most direct evidence that the cue was associated with stated choice in this experiment.

Table 2 reports all observed response sequences and makes the paired analysis reproducible from the aggregate record. No participant had the YYX sequence.

## 4.2 Decision Making: Interview Results

The interviews show why the same renewable-energy cue can lead to different choices. Time was the dominant constraint for P1, P2, P8, and P10, especially when an additional detour threatened work or schedule commitments.

**Table 1: Participant Demographics, Driving Profiles, Region, and Occupation**

| Participant | Gender | Age | Region | Occupation | Driving Time | Daily Battery Usage |
|---|---|---|---|---|---|---|
| P1 | M | 28 | Rural | Commuter | 50 min/day | 25% |
| P2 | M | 29 | Rural | Commuter | 50 min/day | 25% |
| P3 | M | 62 | Urban | Retired older adult | 30 min/day | 10% |
| P4 | F | 61 | Urban | Retired older adult | 15 min/day | 5% |
| P5 | M | 26 | Suburban | Commuter | 120 min/day | 40% |
| P6 | F | 27 | Suburban | Commuter | 120 min/day | 30% |
| P7 | F | 46 | Suburban | E-taxi driver | 600 min/day | 180% |
| P8 | M | 53 | Suburban | E-taxi driver | 600 min/day | 180% |
| P9 | F | 42 | Urban | Commuter | 180 min/day | 60% |
| P10 | F | 25 | Rural | Commuter | 70 min/day | 30% |

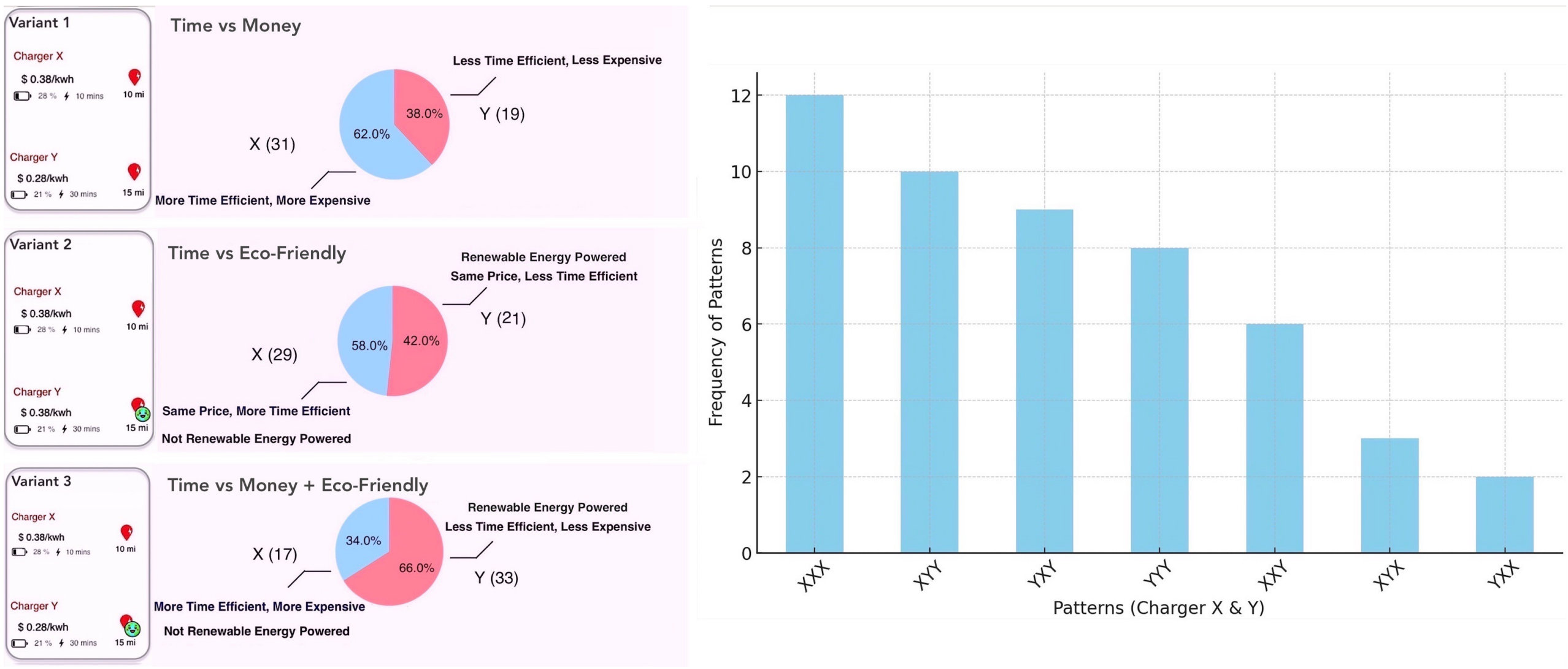

**Figure 3: Aggregate choices across the three interface variants ($N = 50$). X is always the faster station; Y is the slower station. Price and renewable-energy information vary by variant as described in Figure 2.**

**Table 2: Participant-level response patterns across V1, V2, and V3**

| Pattern | $n$ | Share |
|---|---|---|
| XXX | 12 | 24% |
| XXY | 6 | 12% |
| XYX | 3 | 6% |
| XYY | 10 | 20% |
| YXX | 2 | 4% |
| YXY | 9 | 18% |
| YYX | 0 | 0% |
| YYY | 8 | 16% |

> *"My time is more valuable than the money that I save."* (P1)

> *"Depending on how much time it takes to move there (to a renewable-powered station), my answer would differ. 2 minutes are fine; 10 mins are too much."* (P2)

P8, working as a taxi driver, emphasized the economic aspect of time value:

> *"I am likely to make more money for the 10 minutes I waste, so I would definitely not drive extra because it just doesn't make sense to me."* (P8)

Similarly, P10 prioritized efficiency, stating,

> *"I don't care. I have to go to work. I don't want to be stuck in traffic, and also, it's early in the morning when I am really sleepy, or when I am off work, I just head home ASAP."* (P10)

Participants P3, P5, P7, and P9 were more willing to accept additional time when the monetary saving seemed worthwhile.

> *"Definitely the cheaper one even though it's further. It's likely to be on my way already."* (P3)
> *"If it's on my way, I would go to the cheaper one even if it's further. It's just obvious to do that."* (P5)
> *"It's actually almost the same; we are talking about whether I should save a certain amount for this much time. I think it's worth it as I calculate."* (P7)
> *"If I am not in a rush, most likely I want to save some money unless I am in a rush. We are talking about if we have some extra time or not."* (P9)

P4 and P6 explicitly valued renewable energy, but described it together with a monetary benefit rather than as a sufficient reason on its own.

> *"If it can save some money and also protect the environment, I am willing to do so."* (P4) *"It's rewarding me some money for choosing the green earth one. I like doing that because it makes me feel happy."* (P6)

Participants also mentioned weather, traffic, parking or charger availability, nearby amenities, and the station's surroundings. These factors alter the practical meaning of "ten extra minutes" or a small price difference. Several participants said that repeated use led them to rely on memory instead of navigation, which can make a dynamic renewable-energy signal easy to miss. They also anticipated that a recommendation would lose credibility if the suggested station became occupied or if traffic erased the promised benefit. Thus, renewable-energy information competes not only with time and price but also with the reliability of the recommendation itself.

# 5 DISCUSSION

## 5.1 Renewable Energy Matters Within Practical Constraints

The corrected paired analysis sharpens the paper's main finding. When time and price were held in the same relation (V1 vs V3), adding the renewable-energy cue to Y was associated with a 28-percentage-point increase in selection of that slower option. The shift was not universal: 12 participants always selected X, eight always selected Y, and others changed choices as the attributes changed. The interviews explain this heterogeneity. Participants evaluated renewable energy through the opportunity cost of time, the monetary saving, and the risk that real traffic or charger occupancy would invalidate the recommendation.

The V2–V3 comparison provides complementary evidence. Both variants labeled Y as renewable-powered, but making Y cheaper in V3 produced a substantial net movement toward it. This is consistent with prior work showing that price, detour time, convenience, and reliability are central to public-charging choices [40, 63]. More recent choice experiments similarly find a strong role for charging cost and reliability [30], while a study of charging programs estimates that communicating green electricity can offset part of the financial incentive required to shift charging time [54]. Revealed-preference evidence also shows substantial heterogeneity and strong effects of distance and outlet availability [25]. Together, these studies argue against ranking stations with a universal "green" weight.

Figure 1 organizes these findings as a consumer journey [13]. It should be read as a design synthesis of the study, not as evidence that every driver follows the same sequence.

## 5.2 Design Implications

**Use a constraint-first recommendation strategy rather than a universal green score.** A recommender can first remove options that violate a user's maximum detour, arrival deadline, connector requirement, or price ceiling, and then use renewable-energy information to rank the feasible set. The interface should make the trade-off explicit (for example, "8 minutes longer, EUR 2.10 cheaper, higher renewable share") and let the driver change the constraint.

**Communicate provenance and uncertainty.** A binary "green" icon can overstate what is known about electricity supply. Where data permit, the interface should identify whether the signal represents a station's contracted tariff, on-site generation, average grid mix, or marginal carbon intensity; show the applicable time window; and communicate uncertainty or data age. This protects trust and makes the environmental claim auditable.

**Design for reliability and recovery.** Traffic and charger occupancy can erase the benefit that justified a detour. Recommendations should therefore include live availability, a confidence estimate, and a fallback station. If conditions change, the system should explain why it is rerouting rather than silently replacing the recommendation.

**Personalize without locking users in.** The response patterns and interviews suggest stable differences as well as situational shifts. A system may learn whether a driver usually prioritizes time or price, but should expose and allow correction of that profile. Preference learning also raises privacy concerns because location history, schedules, and charging behavior can reveal sensitive routines.

**Minimize in-motion interaction.** Familiar routes and spatial memory can reduce attention to a screen, but proactive prompts may themselves distract. Eco-feedback should be glanceable, defer detailed comparisons until the vehicle is stationary, and use auditory prompts sparingly and in accordance with automotive safety guidance [65].

## 5.3 Limitations and Future Work

This is a small exploratory stated-choice study, not a field evaluation. The quantitative sample includes car users who may be passengers or non-EV owners, whereas the interview sample contains only 10 EV drivers. The archived study materials used for this revision do not provide full quantitative-sample demographics, recruitment and compensation procedures, condition order or counterbalancing, an a priori power analysis, or an ethics/consent statement. These omissions limit reproducibility and should be addressed in future data collection and reporting.

The three scenarios do not form a complete factorial design. V1–V3 supports a focused comparison because the time–price relation is preserved, but the renewable-energy cue remains tied to option Y and therefore may be confounded with position or sequence. V1–V2 changes two attributes simultaneously. A future study should randomize order and position, manipulate time, price, renewable share, occupancy, and reliability orthogonally, and use a discrete-choice model with repeated observations per participant [16]. The study also uses a binary renewable-energy label rather than a verified percentage or carbon-intensity forecast.

Only aggregate response-pattern counts were available for the corrected analysis, so we could not estimate heterogeneity by participant characteristics or attach choices to interview accounts. The interview report does not document a formal coding protocol. Future work should publish de-identified choice data, materials, code, and a qualitative codebook where consent permits. Most importantly, deployment studies should compare stated choices with actual navigation and charging behavior over time, including failures, rerouting, trust, and safety.

# 6 CONCLUSIONS

Renewable-energy information can matter in EV charging recommendations, but it is interpreted through time, price, and confidence that the station will remain a viable option. In the clearest within-person comparison, adding a renewable-energy cue while preserving the time–price trade-off increased selection of the slower station from 38% to 66%. Interviews showed why this shift was conditional: urgency, income, traffic, charger occupancy, nearby amenities, and familiarity changed the perceived cost of acting on the recommendation. The study therefore offers both an empirical baseline for renewable-energy-aware EV charging recommendations and a constraint-first design rationale: rank sustainable options within user constraints, explain the basis and uncertainty of the environmental signal, and offer reliable fallbacks. Larger randomized and longitudinal field studies are needed before these stated-choice results can be generalized to real charging behavior.

*This revised draft is an updated version of the paper titled*

*"Time and Money Matters for Sustainability: Insights on User Preferences on Renewable Energy on Electric Vehicle Charging Stations'*

# ACKNOWLEDGMENTS

This project has received funding from the European Union's Horizon 2020 research and innovation programme under the Marie Skłodowska-Curie grant agreement No 955422.

We extend our appreciation to the following individuals for their feedback and support: Daniel Korus, Danielle Geng, Dennis Lawo, Lihua Kong, Lukas Böhm, Guoqing Wang, Jingjing Xu, Kai Mannwald, MD Shajalal, Sidra Naveed, and Yinping Du; we also want to thank all the study participants and fellow researchers and supervisors of the Gecko project [15].